\documentclass[11pt,a4paper]{article}
\usepackage{jcappub}

\usepackage{lipsum}

\usepackage{graphicx}
\usepackage{dcolumn}
\usepackage{bm}
\usepackage{hyperref}
\usepackage{xcolor}
\usepackage{float}
\usepackage{dsfont}

\usepackage{import}
\usepackage{xifthen}
\usepackage{pdfpages}
\usepackage{transparent}
\usepackage{comment}
\usepackage{cancel}
\usepackage{sidecap}

\newcommand{\ket}[1]{\left\lvert#1\rangle\right.}

\title{Polymer Geodesic Motion in Schwarzschild Spacetime}

\author[a]{Lorenzo Boldorini,}
\author[a]{Corrado Marzano,}
\author[b,a]{Giovanni Montani}

\affiliation[a]{Department of Physics, “La Sapienza” University of Rome,\\ P.le Aldo Moro 5, 00185 Rome, Italy}
\affiliation[b]{ENEA, Fusion and Nuclear Safety Department, C. R. Frascati,\\Via E. Fermi 45, 00044 Frascati (RM), Italy}

\emailAdd{boldorini.1843532@studenti.uniroma1.it}
\emailAdd{marzano.1918552@studenti.uniroma1.it}
\emailAdd{giovanni.montani@enea.it}

\abstract{
In this paper we will study the geodesic motion of massive particles, in a Schwarzschild background, with a semi-classical quantum framework called "Polymer Quantum Mechanics" (PQM) in order to investigate the black hole phenomenology resulting from this formulation, which accounts for Planckian scale physics. In particular we studied two main scenarios, being the radial in-fall and circular orbits and their stability. In this framework, we built an effective Hamiltonian taking into account the polymer quantum effects, altering the classical equations of motion with Planckian scale corrections. As a main result, we obtained the existence of a classically forbidden region surrounding the event horizon, preventing particles from crossing it. Additionally, we discovered the presence of stable circular orbits below both the classical Innermost Stable Circular Orbit (ISCO) and the horizon (corresponding to closed time-like geodesics).
}

\begin{document}

\date{\today}

\maketitle

\section{Introduction}
\label{sec:intro}
The idea that the physical space 
has a discrete nature on a Planckian scale is a well-established conjecture in canonical quantum gravity \cite{Thiemann_2007} and also 
in other quantum formulation of 
gravity \cite{KieferREV}.
Surely, the most relevant physical 
insight, obtained till now, on the 
physical space discreteness comes 
from Loop Quantum Gravity \cite{Rovelli_2004} and, in particular, from 
the emergence of a discrete spectrum
characterizing, on a kinematical level, 
geometrical operators like areas and 
volumes \cite{Rovelli_1994,Rovelli_1995}.\\
In different approaches, like String 
Theories \cite{Polchinski1998-yz,Polchinski2005-ac} or 
non-commutative spacetime formulations 
\cite{Seiberg:1999vs,Amelino-Camelia}, the implications 
of a discrete-like nature of the Planckian scale are mainly transferred to 
quantum particle dynamics. 
By other words, highly energetic particles approaching a Planckian energy scale, are able to experience the 
physical scale discreteness and, hence, 
their dynamical properties are 
correspondingly altered.\\
This question is of particular relevance when we face the study of Black Hole Physics \cite{Misner:1973prb}, since the 
nature of such astrophysical objects is mainly manifested via the behavior of free falling (massive or massless) particles in their gravitational field. In this respect, having in mind to 
explore implications of the free falling particles in a Black Hole, coming from the Planckian scale physics, 
we are naturally lead to require 
the diffeomorphism invariance of the 
required quantum dynamical scenario. 
As well-known \cite{Morchio_2007, Strocchi2016}, addressing such a request on a generic phase space formulation, implies the impossibility to properly define one of the two 
conjugated variables. Then, 
in order to make the obtained quantum theory technically viable in the presence of such ill-defined operators, a lattice structure is introduced for one of the two variables, making possible to approximate the ill-defined operator with a difference one\cite{Corichi2007}.
This approach is dubbed "Polymer Quantum Mechanics" (PQM) and for a semi-classical implementation to the gravitational collapse, see \cite{Barca2023, Boldorini_2024}.\\
When dealing with point-like free falling particles in a Black Hole, 
it is natural to assume as a discrete-like coordinate the radial position, 
since spherical symmetry confers 
a privileged character to it. 
As a result, the conjugate momentum to the particle radial coordinate is 
replaced in the Hamiltonian, according to the PQM formulation, by a sine function of itself, also depending on the 
characteristic Planckian-like scale of 
the radial discretization. 
While such a reformulation would mainly concern the quantization of the 
particle dynamics in the momentum representation, according to the Ehrenfest conjecture quantum mean values 
follow a classical trajectory when the 
state is properly localized, i.e. it has quasi-classical features. 
It is of relevant interest the study 
of a classical polymer particle, 
which Hamiltonian has now an effective character.\\
It is exactly the paradigm of a 
semi-classical polymer particle, free falling in a classical Schwarzschild 
geometry, that we address here. 
The resulting physical insight we 
can infer from our set-up is 
a qualitative understanding of 
how the discrete nature of the 
physical scale, here translated into 
a discrete radial coordinate of the 
particle, can influence the Black Hole phenomenology, i.e. the role of 
a Schwarzshild horizon, the morphology 
of stable circular orbits etc..
The proposed revised dynamical picture will relay on the modified nature 
that the effective potential takes 
for a polymer particle in the Schwarzshild spacetime, from which all the 
main dynamical consequences naturally 
follow.\\
We stress how, when studying extreme 
astrophysical processes, it is 
immediate to deal with the dynamics 
of highly energetic fundamental particles, which quantum nature is described via a quasi-classical representation. It is just this consideration the 
most important motivation for the 
present study and, at the same time, 
a reliable justification for the 
set-up of our point-particle model.\\
More specifically, we are interested in the geodesic motion of massive particles in the classical Schwarzschild geometry, in order to understand what features this model introduces to classical geodesics and whether or not such features are consistent with our current observations.\\
Our main findings show the appearance of an impermeable surface above the event horizon, where the particles bounce off from, and the presence of stable circular orbits below the classical Innermost Stable Circular Orbit.\\
The manuscript is organized as follows. In section \ref{sec:classical schwarzschild} we give a brief overview of the geometrical setup behind geodesic motion in the classical Schwarzschild geometry, in particular we focus on the radial capture of massive particles and their circular orbits.\\
In section \ref{sec:hamil_pol} we introduce the Polymer Quantization framework as a way of quantizing particles in a diffeomorphism invariant manner, best suited for description of quantum particles on manifolds.\\
Section \ref{sec:rad_pol} is focused on the study of those polymer particles when they are radially in-falling, characterizing the new features that emerge in the newly introduced framework. The most surprising result is that the horizon is classically impermeable, and if one does not consider tunneling properties particles always bounce back when approaching the event horizon.\\
Section \ref{sec:circ_pol} analyzes the presence of circular orbits in the new quantization scheme, with the main result ending up with the presence of stable circular orbits in a region that classically prevents them.\\
At last section \ref{sec:concl} closes this paper with a summary of the obtained results and an overview on the future perspectives that could be further studied.\\
In this work we adopted the signature convention for the metric tensor $Sign(g) = (-,+,+,+)$ and the natural units given by $G=c=\hbar=1$.
\section{Classical Schwarzschild Massive Geodesics}
\label{sec:classical schwarzschild}
\begin{figure}[H]
	\centering
	\includegraphics[width=0.69\linewidth]{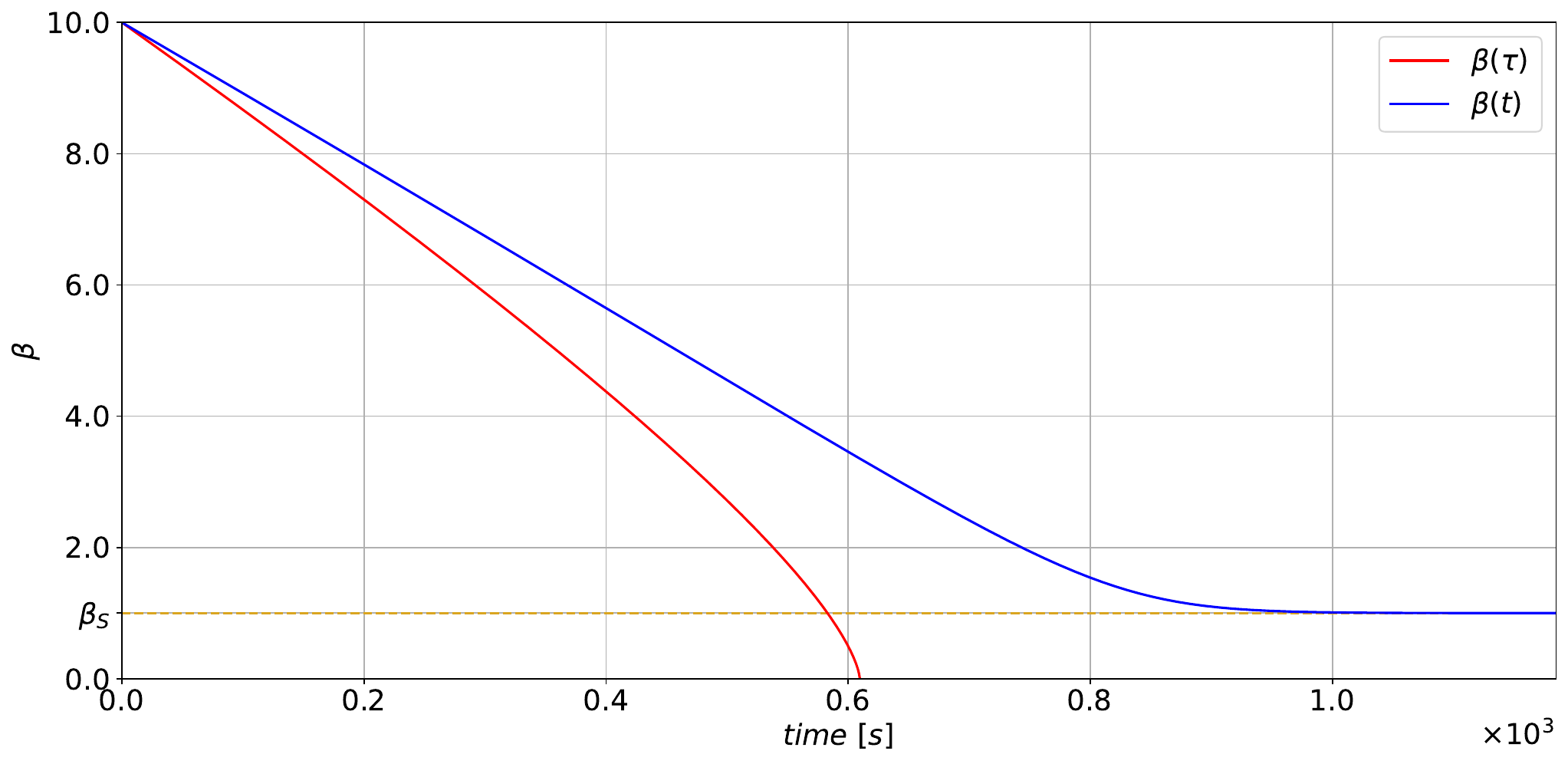}
	\caption{Proper time $\tau$, in red, and coordinate time $t$, in blue, of a massive particle in-falling radially from the initial position $\beta_0$ into a black hole.}
	\label{fig:t&tau(r) classic}
\end{figure}

\begin{figure}[H]
  \centering
  \begin{minipage}[b]{0.49\textwidth}
    \centering
    \includegraphics[width=\textwidth]{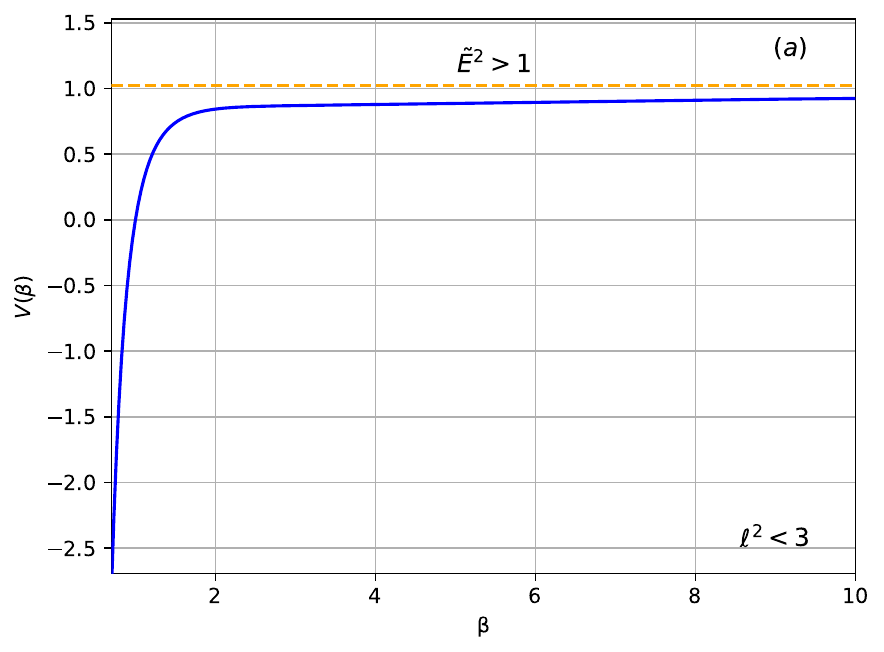}
  \end{minipage}
  \begin{minipage}[b]{0.49\textwidth}
    \centering
    \includegraphics[width=\textwidth]{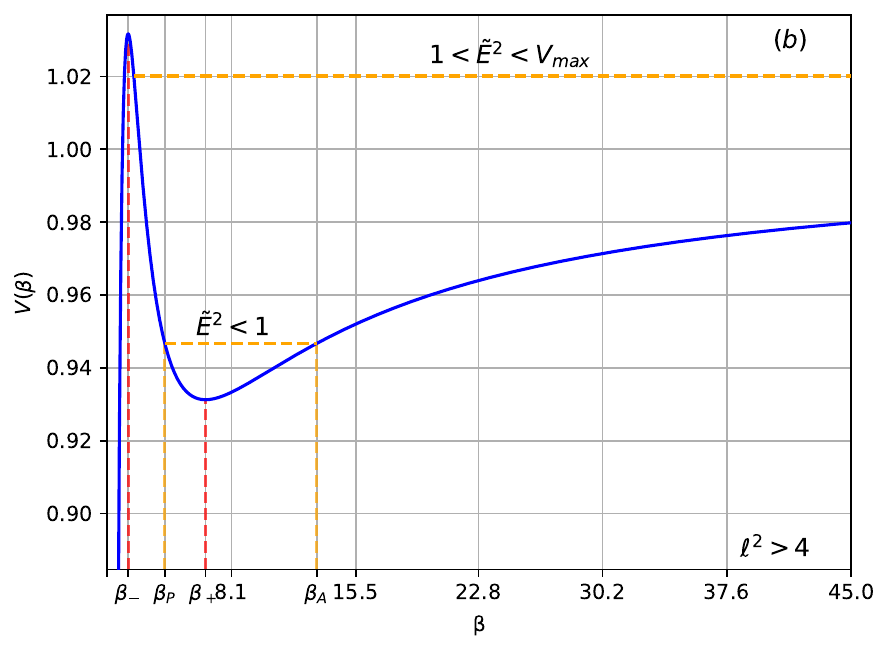}
  \end{minipage}
  \begin{minipage}[b]{0.49\textwidth}
    \centering
    \includegraphics[width=\textwidth]{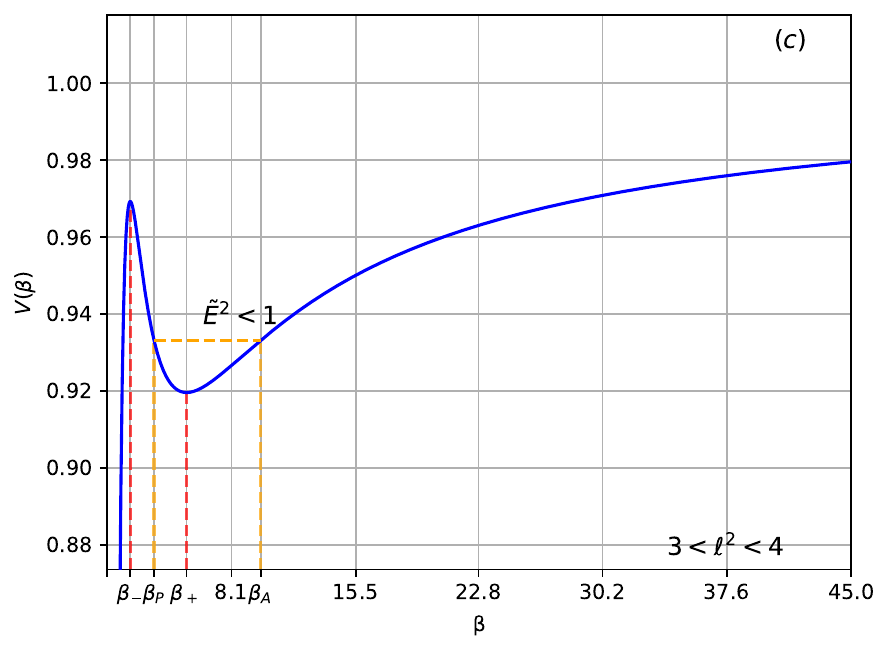}
  \end{minipage}

  \caption{Effective potential (blue) for a massive particle in some representative cases. In red are highlighted the circular orbits radii $\beta_\pm$ and in orange the periastron $\beta_P$ and apastron $\beta_A$ associated to a given energy $\tilde E^2$.}
  \label{fig:V(r) classic}
\end{figure}

We begin our analysis by recalling the classical Schwarzschild\cite{Pani} setup, and we will work with the following variables to have a description less dependent on the black hole and the test particle specific quantities:
\begin{equation}
    \beta = \frac{r}{r_S} \qquad \tilde E = \frac{E}{m} \qquad \ell = \frac{L}{m \ r_S}
\end{equation}
Where $m$, $E$ and $L$ are respectively the test particle's mass, energy and angular momentum.\\
We report the classical static and spherically symmetric Schwarzschild metric in those variables, in Schwarzschild coordinates:
\begin{equation}
		ds^2 = -\left(1-\frac{1}{\beta}\right) dt^2 + \frac{r_S^2 \ d\beta^2}{1-\frac{1}{\beta}} + r_S^2 \ \beta^2 (d\theta^2 + \sin^2(\theta) d\varphi^2)
	\label{eq:schwarzschild metric}
\end{equation}
From which it is easy to obtain, together with the planarity property of the particle motion in a Schwarzschild background, the Equations of Motion through the Geodesic Equations, given by:
\begin{align}
    \dot t = -\frac{\tilde E}{1-\frac{1}{\beta}}\qquad
    r_S^2 \ \dot \beta^2 = \tilde E^2 - V(\beta) \qquad
    \dot \theta = 0 \qquad
    r_S \dot \varphi = \frac{\ell}{\beta^2}
    \label{eq:classical geodesics}
\end{align}
Where the dot quantities represent the derivatives with respect to the proper time $\tau$.
\\
The function $V(\beta)$ is called the Effective Potential and for massive particles has the form:
\begin{equation}
    V(\beta) = \left(1-\frac{1}{\beta}\right)\left(1+\frac{\ell^2}{\beta^2}\right)
    \label{eq:classic V(r)}
\end{equation}
Which delineates a forbidden region in the $\tilde E^2 \ vs \ \beta$ plane, where $\tilde E^2 < V(\beta)$ implies $\dot \beta^2 < 0$.\\
When $V(\beta) = \tilde E^2$ the particle reaches an inversion point, where it either bounces off the potential or stays in that position to not trespass in the forbidden region: the smaller value, \( \beta_P \), is the periastron, the point of minimum distance from the central body. The largest, \( \beta_A \), is the apastron, the point of maximum distance.\\
We initially consider the radial capture of a massive particle falling towards a black hole, so $L = 0$, $\frac{d \varphi}{d \tau} = 0$ and $\dot \beta < 0$.\\
Defining the initial position of the motion as $\beta_0 \equiv \beta(\tau = 0)$, we can find the expression for $\tau$ by inverting and integrating the $\dot \beta$ equation, and the one for $t$ by dividing and integrating the $\dot \beta$ and $\dot t$ equations:
\begin{equation}
      \tau(\beta) = - \int_{\beta_0}^\beta \ \frac{r_S \ d\beta'}{\sqrt{\tilde E^2 - 1 + \frac{1}{\beta'}}}\qquad
        t(\beta) =  \int_{0}^t dt' = - \int_{\beta_0}^\beta \frac{\tilde E}{1-\frac{1}{\beta'}} \ \frac{r_S \ d\beta'}{\sqrt{\tilde E^2 - 1 + \frac{1}{\beta'}}}
	\label{eq:t(r)&tau(r) classic}
\end{equation}
It can be clearly seen that while the proper time $\tau$ is regular on the horizon $t$ diverges on $\beta=1$, meaning that an external observer will never see the particle crossing the horizon (see figure \ref{fig:t&tau(r) classic}).
We are interested also in the study of circular orbits. To derive them, we analyze the potential to find whether it admits a minimum or a maximum, needing to solve the equation $V'(\beta) = 0$, which has the classical solutions:
\begin{equation}
	\beta^\pm_{cl} = \ell^2 \pm \sqrt{\ell^4 - 3 \ell^2} 
	\label{eq:r pm classic}
\end{equation}
In figure \ref{fig:V(r) classic} we report three cases of the effective potential considering three different particle's angular momentum regimes generated by these roots.\\
When $\ell^2 < 3$, the roots do not exist and so there are no inversion points for a particle coming from infinity. Instead if $3 < \ell^2 < 4$, the roots exist and $V_{max} \equiv V(\beta_-) < 1$. At last for $\ell^2 > 4$, the roots exist and $V_{max} \equiv V(\beta_-) > 1$.
Then studying circular orbits determined by the extremal points of the effective potential, where stable orbits are identified by the minimums $V(\beta_+) \equiv V_{min}$ and unstable ones by the maximums $V(\beta_-) \equiv V_{max}$, it can be demonstrated that the Innermost Stable Circular Orbit (ISCO) is positioned at $\beta_{ISCO} = 3$, while the Innermost Unstable Circular Orbit (IUCO) is positioned at $\beta_{IUCO} = \frac{3}{2}$.\\

\section{The Effective Hamiltonian Model}
\label{sec:hamil_pol}
In this section first we will analyze a quantization procedure suitable for quantum particles on classical manifolds and then we will see how to apply it to a particle free-falling on said manifold.\\
Since the main principle behind General Relativity lies in the fact that coordinates choices are not physical, so that the laws of physics must be diffeomorphism-invariant, one might want to quantize particles in a diffeomorphism invariant way.\\
The proposed solution is that to build a diffeomorphism invariant ground state via the Gelfand-Naimark-Segal (GNS) construction\cite{Segal1947v1,Segal1947Irreducible,GelfandNaimark1943}, implementing the quantization via a non-regular Weyl scheme\cite{Acerbi:1992yv}. \\
We start by recalling the description of diffeomorphisms introduced in \cite{Morchio_2007} and later expanded in \cite{Strocchi2016}.
Given $C^\infty(\mathcal{M})$, the space of smooth functions on $\mathcal{M}$, and $\text{Vect}(\mathcal{M})$, the Lie algebra of $C^\infty$ vector fields on $\mathcal{M}$ of compact support, we introduce the Weyl operators for $f\in C^\infty(\mathcal{M})$ and $\bm{v} \in  \text{Vect}(\mathcal{M})$:
\begin{equation}
\label{eqn:diffeo_weyl}
\begin{cases}
W(f) = e^{if} \qquad f \in  C^\infty(\mathcal{M})\\
V(\bm{v}) = e^{i\bm{v}} \qquad \bm{v} \in \text{Vect}(\mathcal{M})
\end{cases}
\end{equation}
They generated the polynomial Weyl algebra $\mathcal{A}$ that is a $^*-$algebra, where $\text{Diff}(\mathcal{M})$ defines a group of $^*-$automorphisms of $\mathcal{A}$, denoted by $\alpha_h$ for $h\in \text{Diff}(\mathcal{M})$:
\begin{equation}
\alpha_h\left(W(f)\right) = W(f_h) \qquad \alpha_h\left(V(\bm{v})\right)  = V(h\bm{v})
\end{equation}
On a manifold, a quantum particle regular representation $\pi$ of the $^*-$algebra $\mathcal{A}$, generated by the Weyl operators (\ref{eqn:diffeo_weyl}), is a $^*-$homomorphism of  $\mathcal{A}$ in the $^*-$algebra of bounded operators on an Hilbert space $\mathcal{H}_\pi$.\\
We can than define a Hilbert space representation defining a diffeomorphism-covariant quantum system.\\
Such particle representation is locally unitary equivalent to the standard Schrödinger one in $L^2(\mathcal{M}, d\mu)$ with $W(f)$ acting as a multiplication operator and $V(\bm{v})$ acting on $\Psi \in L^2(\mathcal{M}, d\mu)$ as:
\begin{equation}
\label{eqn:diffeo_quantum_V}
\pi\left(V(\bm{v})\right)\Psi(x) = \Psi\left(g(\bm{v})^{-1}(x)\right) \sqrt{\frac{d\mu\left(g(\bm{v})^-1(x)\right)}{d\mu(x)}}
\end{equation}
Where $g(\bm{v})\in \text{Diff}(\mathcal{M})$ is the group element associated to $\bm{v}$.\\
A state is diffeomorphism invariant if:
\begin{equation}
\omega(\alpha_{\bm{v}}(A)) = \omega(A) \qquad \forall A \in \mathcal{A} , \forall \bm{v} \in \text{Vect}(\mathcal{M})
\end{equation}
Thanks to the GNS construction we know that $\text{Diff}(\mathcal{M})$ is implemented by the unitary operators $U(\lambda\bm{v})$ defined as:
\begin{equation}
U(\lambda\bm{v})A\Psi_\omega = \alpha_{\lambda\bm{v}}(A)\Psi_\omega
\end{equation}
This implies that:
\begin{align}
\omega\left(W(f)V(\bm{v})\right) =\omega\left(W(f)\right) = & \;\omega\left(W(g(\bm{w})(f))\right) \\ 
\forall f \in  C^\infty(\mathcal{M})\qquad & \forall \bm{v},\bm{w} \in \text{Vect}(\mathcal{M})
\end{align}
We now see that the GNS representation defined by a diffeomorphism invariant state must be a non-regular quantum particle representation.
Diffeomorphism invariance tells us that:
\begin{equation}
\omega(\left[\bm{v}, W(f)\right]) = 0
\end{equation}
This is incompatible with equation (\ref{eqn:diffeo_quantum_V}), and so with the regularity hypothesis.\\
We see that a GNS representation $\pi$ such that the diffeomorphism invariant state $\omega$ with $\omega(W(f)) = 0 \;\;\forall f \neq const \ \mathds{1} $ is characterized by the fact that the $\pi(V(g(\lambda\bm{v})))$ are not weakly continuous, so the corresponding generators do not exist as operators in $\mathcal{H}_\omega$, and two states $W(f)\Psi_\omega$ and $W(g)\Psi_\omega$ are orthogonal unless $f-g = const \hspace{0.5mm} \mathds{1}$.
This are exactly the polymer states with a ill-defined $\hat{p}$ operator introduced in \cite{Corichi2007}.\\
The above exposition shows that this representation has a discrete continuous set of coordinates, and that the momentum operator is not well defined quantumly.\\ 
When dealing with an Hamiltonian of the kind:
\begin{equation}
H(q,p) = K(p) + V(q)
\end{equation}
The operator corresponding to the kinetic term needs to be approximated. The solution is to work with a lattice on $q$ with step $\mu_0$, which needs to be refined.\\
In such a scenario the shift operator shifts the states by a discrete amount, so that:
\begin{equation}
\hat{T}(\mu_0)\ket{\mu} = e^{i\mu_0 p}\ket{\mu} =\ket{\mu+\mu_0}
\end{equation}
With this we can use the approximate:
\begin{equation}
\hat{p}\simeq \frac{1}{\mu_0} \frac{1}{2i}(\hat{T}(\mu_0)-\hat{T}(-\mu_0))\simeq \frac{1}{\mu_0} \frac{1}{2i}(e^{i\mu_0 p} - e^{-i\mu_0 p})
\end{equation}
This gives an approximation for $\hat{p}$ in the limit $\mu_0\rightarrow 0$:
\begin{equation}
\hat{p}\psi(p) = \frac{1}{\mu_0} \sin\left(\mu_0 p\right)\psi(p) 
\end{equation}
Such regularization procedure is also useful to define effective Hamiltonians, where the effective corrections are given by:
\begin{equation}
\label{eqn:polymer_corr}
p^2 \rightarrow \frac{1}{\mu_0^2} \sin\left(\mu_0 p\right)^2
\end{equation}
The action for the particle, invariant  under world-line reparametrizations, is given by:
\begin{equation}
	S[x^\mu, p_\mu, N] = \int d\lambda \left[p_\mu \dot{x}^\mu - \underbrace{\frac{N}{2}\left(g^{\mu\nu} p_\mu p_\nu + m^2\right)}_{H}\right]
	\label{eq:action2}
\end{equation}
Where $N = \frac{d\tau}{d\lambda}$ is  the lapse function, and its equation of motion yields the Hamiltonian constraint $g^{\mu\nu} p_\mu p_\nu + m^2 = 0$.\\
We will work with $N=1/m$, equivalent to the proper time parametrization of $S$.\\
When the planarity conditions $p_\theta=0$ and $\theta=\frac{\pi}{2}$ are satisfied, recalling that $N=1/m$, $p_t = -E$ and $p_\varphi = L$, will yield the following Hamiltonian:
\begin{equation}
    	H = \frac{m}{2}\left(1 -\frac{\tilde E^2}{1-\frac{1}{\beta}}+\frac{p_r^2}{m^2}\left( 1-\frac{1}{\beta} \right) + \frac{\ell^2}{\beta^2}\right)
\end{equation}
The polymer corrections in equation (\ref{eqn:polymer_corr}) will then be inserted into $H$, giving an effective Hamiltonian given by:
\begin{equation}
\label{eqn:full_ham}
        	H_{poly} = \frac{m}{2}\left(1 -\frac{\tilde E^2}{1-\frac{1}{\beta}}+\frac{sin^2(\mu_0 p_r)}{\alpha^2}\left( 1-\frac{1}{\beta} \right) + \frac{\ell^2}{\beta^2}\right)
\end{equation}
Where we defined:
\begin{equation}
    \alpha = m \mu_0
\end{equation}
With this effective Hamiltonian we are now ready to start our analysis of the particle dynamics.
\section{Purely Radial Geodesic} 
\label{sec:rad_pol}
We start by analyzing the case of a massive particle in-falling radially towards a Schwarzschild black hole, so we set the angular momentum $L$ equal to $0$ in the Hamiltonian of equation (\ref{eqn:full_ham}), so  the Hamiltonian constraint becomes:
\begin{equation}
	1 -\frac{\tilde E^2}{1-\frac{1}{\beta}} + \frac{\sin^2(\mu_0 p_r)}{\alpha^2}\left( 1-\frac{1}{\beta} \right)  = 0
	\label{eq:polymer radial constraint}
\end{equation}
The Hamilton equations derived from the radial polymer Hamiltonian, where there is no angular motion, become instead:
\begin{equation}
	\begin{cases}
		\dot{t} = \frac{\tilde E}{1-\frac{1}{\beta}} \\
		\dot{\beta} = \frac{\left(1-\frac{1}{\beta}\right)}{\alpha r_S} \sin (\mu_0 p_r) \cos (\mu_0 p_r) \\
	\end{cases}
	\quad ; \quad
	\begin{cases}
		\dot E = 0 \\
		\dot{p}_r = -\frac{m}{2 \ r_S} \left(\frac{\sin ^2(\mu_0 p_r)}{\alpha^2 \beta^2}+\frac{\tilde E^2 }{\beta^2 \left(1-\frac{1}{\beta}\right)^2}\right) \\
	\end{cases}
	\label{eq:hamilton radial equations var&mom}
\end{equation}
The first thing that we are interested in is to understand whether or not the in-falling particle bounces off at a finite radius. This is achieved by studying $\dot{\beta}^2 =0$ to find the inversion points of the dynamic.\\
To do so we need first to remove from the equation the dependence on $p_r$, so in order to achieve this result we isolate the sine squared from the Hamiltonian constraint in equation (\ref{eq:polymer radial constraint}), obtaining:
\begin{equation}
    \sin^2(\mu_0 p_r) = \frac{1}{1-\frac{1}{\beta}}\alpha^2 \left(\frac{\tilde E^2}{1-\frac{1}{\beta}}-1\right)
\end{equation}
Substituting this into $\dot{\beta}^2$ yields:
\begin{equation}
	\left( \frac{\beta\tilde E^2}{\beta-1} -1 \right)\left[ \left(1-\frac{1}{\beta} \right) + \alpha^2 \left( 1 -\frac{\beta\tilde E^2}{\beta-1} \right) \right] = 0
	\label{eq:rdot2=0 polymer dimensionless}
\end{equation}
This equation has four solutions:
\begin{equation}
		\beta_1 = \frac{1}{1 - \tilde E^2}\qquad
		\beta_2^0 =\frac{1}{2+\alpha^2}\qquad
		\beta_2^\pm = \frac{\alpha^2 +2 \pm \alpha \sqrt{4\tilde E^2 + \alpha^2}}{2\alpha^2(1-\tilde E^2) + 2}
	\label{eq:rdot2=0 polymer dimensionless solutions}
\end{equation} 
We now study the behavior of this points, with the help of a quantity, given by $\tilde E_{crit}^2 \equiv \frac{1+\alpha^2}{\alpha^2}$, referred as the critical energy.\\
The solution $\beta_1$ is always negative or does not exist for $\tilde{E}^2 \geq 1$ and in this regime can be discarded.
When $\tilde{E}^2 < 1$, it is always positive, so:
	\begin{align}
		 \tilde{E}^2 \geq& 1 \qquad\Longrightarrow \qquad \beta_1 < 0 \text{ or not defined,}\\
	    \tilde{E}^2 <& 1\qquad \Longrightarrow\qquad \beta_1 > 1 
		\label{eq:beta1 study}
	\end{align}
The solution $\beta_2^0$ holds only if $\tilde{E}^2 = \tilde E_{crit}^2 \equiv \frac{1+\alpha^2}{\alpha^2}$, which is only compatible with the negative $\beta_1$ solution.
The solutions $\beta_2^\pm$ are valid everytime that $\tilde{E}^2$ is both greater or equal to the critical value $\tilde E_{crit}^2$. The condition $\tilde{E}^2 > \tilde E_{crit}^2$ is only compatible with the negative $\beta_1$ solution, while the condition $\tilde{E}^2 < \tilde E_{crit}^2$ is compatible with both the $\beta_1$ solutions.\\
So, to summarize, the values of the solutions are distributed as follows:
\begin{align}
	 &\tilde{E}^2= \tilde E_{crit}^2 \equiv \frac{1+\alpha^2}{\alpha^2}\qquad &\Longrightarrow\qquad& \beta_1<0<\beta_2^0<\frac{1}{2}\\
	 &\tilde{E}^2 > \tilde E_{crit}^2\qquad &\Longrightarrow\qquad& \beta_2^+<\beta_1<0<\beta_2^-<1\\
	&1\leq\tilde{E}^2 < \tilde E_{crit}^2 \qquad&\Longrightarrow \qquad&\beta_1<0<\beta_2^-<1<\beta_2^+ \\
	 &\tilde{E}^2 < 1< \tilde E_{crit}^2 \qquad&\Longrightarrow \qquad& 0<\beta_2^-<1<\beta_2^+ <\beta_1
	\label{eq:beta|rdot2=0 polymer radial study}
\end{align}
This means that $\beta_2^+$ can be discarded when the condition $\tilde{E}^2 > \frac{1+\alpha^2}{\alpha^2}$ holds and that $\beta_1$ can be discarded when the condition $\tilde{E}^2 \geq 1$ holds. Notice that $\beta_1$ approaches infinity as $\tilde{E}^2$ approaches 1 from below.
\begin{figure}[H]
	\centering
	\def\svgscale{0.58}
    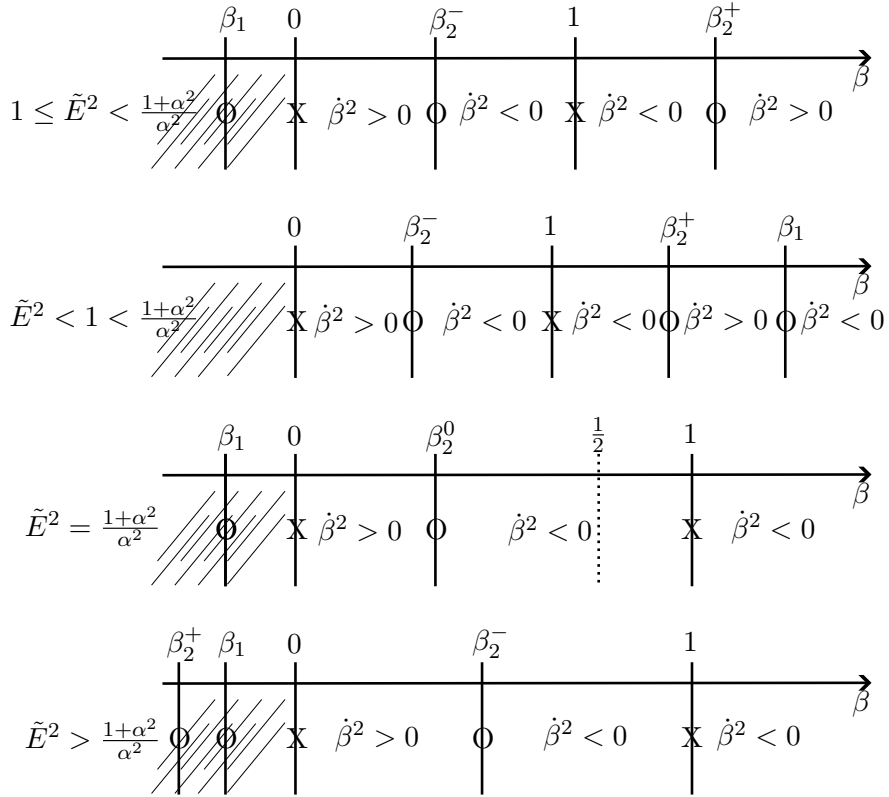

	\caption{Study of the sign of the squared velocity in the four cases.}
	\label{fig:sds_radial}
\end{figure}
To understand those results we compare them with the sign of $\dot{\beta}^2$, shown in figure \ref{fig:sds_radial}.\\
It is clear that the condition that $\dot{\beta}^2>0$ outside the event horizon is satisfied only when $\tilde{E}^2 < \frac{1+\alpha^2}{\alpha^2}$, so the other solutions must be ignored, being non-physical. The two conditions $\tilde{E}^2 \geq 1$ and $\tilde{E}^2 < 1$ define respectively unbounded and bounded orbits, so that $\beta_2^+ \equiv \beta_{min}$ is the minimal reachable radius that defines the inversion point for unbounded orbits (the periastron).\\
$\beta_1 \equiv \beta_{max}$ is, instead, the maximal distance reachable with a bounded orbit, and a radially moving particle must oscillate between $\beta_{min}$ and $\beta_{max}$.\\
So the solutions we found and their respective ranges of validity will be the following:
\begin{align}
		 1\leq\tilde{E}^2 < \tilde E_{crit}^2 \qquad&\Longrightarrow \qquad \beta_2^+>1 \\
	 \tilde{E}^2 < 1< \tilde E_{crit}^2 \qquad&\Longrightarrow \qquad 1<\beta_2^+ <\beta_1
	\label{eq:beta|rdot2=0 polymer radial study outside horizon}
\end{align}
Notice that in the limit of low energies, when $\tilde{E}^2 << \tilde E_{crit}^2$, and Planckian lattice step, $\mu_0 \approx \ell_P$,  the new inversion point $\beta_2^+$ approaches the Schwarzschild radius at $\beta_S= 1$.\\
Under the same conditions, but with $E < 1$, $\beta_1$ always exists out of the horizon and the particle, which cannot come from infinity due to being in a bounded orbit like in the classical case, is trapped in a closed trajectory between $\beta_2^+$ and $\beta_1$.\\
Furthermore since $\tilde E_{crit}^2 = 1 + \frac{1}{\mu_0^2 m^2}$ if $\mu_0 \approx \ell_P$ then the value of $\tilde E_{crit}$ is of the order of the Planck energy.\\
Next we evaluated the radial time geodesics, both in proper and coordinate time, which we obtained from the equations (\ref{eq:hamilton radial equations var&mom}), following the same procedure described in section \ref{sec:classical schwarzschild}:
\begin{align}
    &\tau(\beta) = \int_{\beta_{0}}^{\beta} \pm \frac{d\beta' \ r_S}{ \sqrt{\left( -1 +\frac{\beta'\tilde E^2}{\beta'-1} \right)\left[ \left(1-\frac{1}{\beta'} \right) + \alpha^2 \left( 1 -\frac{\beta'\tilde E^2}{\beta'-1}  \right) \right]}} 
    \label{eq:tau polymer}
    \\ 
    &t(\beta) = \int_{\beta_{0}}^{\beta} \pm \frac{\left(\tilde{E} \ r_S\right)\beta' d\beta'}{\left(\beta' -1\right)\sqrt{\left( -1 +\frac{\beta'\tilde E^2}{\beta'-1} \right)\left[ \left(1-\frac{1}{\beta'} \right) + \alpha^2 \left( 1 -\frac{\beta'\tilde E^2}{\beta'-1}  \right) \right]}}
    \label{eq:t polymer}
\end{align}
Firstly we integrated the in-falling (negative) equation from a starting point $\beta_0$ up to the inversion point $\beta_{inv}$, then  to evaluate the bounce on the inversion point we integrated the outgoing (positive) equation from the previously reached inversion point $\beta_{inv}$ up to the starting point $\beta_{0}$.\\
The previous geodetic integrals are not analytically solvable, so we numerically studied three different cases for a proton in-falling on $Sgr \ A^*$ \cite{SgrA*1,SgrA*2,SgrA*3,SgrA*4,SgrA*5,SgrA*6,SgrA*7} in different energy regimes with $\mu_0$ at the Planck scale, to verify and evaluate the behavior of the polymer representation's bounce on the inversion point.\\
In the first case we analyzed the energy regime corresponding to the cosmic ray with the highest energy ever measured, the \textit{"Oh-My-God particle"} \cite{OMG_particle}, in order to study the most extremely energetic observations and how the polymer representation modifies and interprets such results, where the polymer is still dominated by the classical Schwarzschild terms because $E_{OMG} \ll E_P$.\\
Later we studied the ultra-high energy regime where the polymer terms are strongly dominant, given by $E = 0.5 \hspace{1mm} E_P$, to better understand the polymer particle behavior, significantly different from the classical case.\\
Lastly, we examined a bound trajectory with $\tilde E  < 1$, realized by $E = 0.5 \hspace{1mm} GeV$, representing a particle trapped bouncing between two inversion points, which is a completely unique scenario with no counterpart in the classical case.
\begin{figure}[H]
    \centering
    \includegraphics[width=1.0\textwidth]{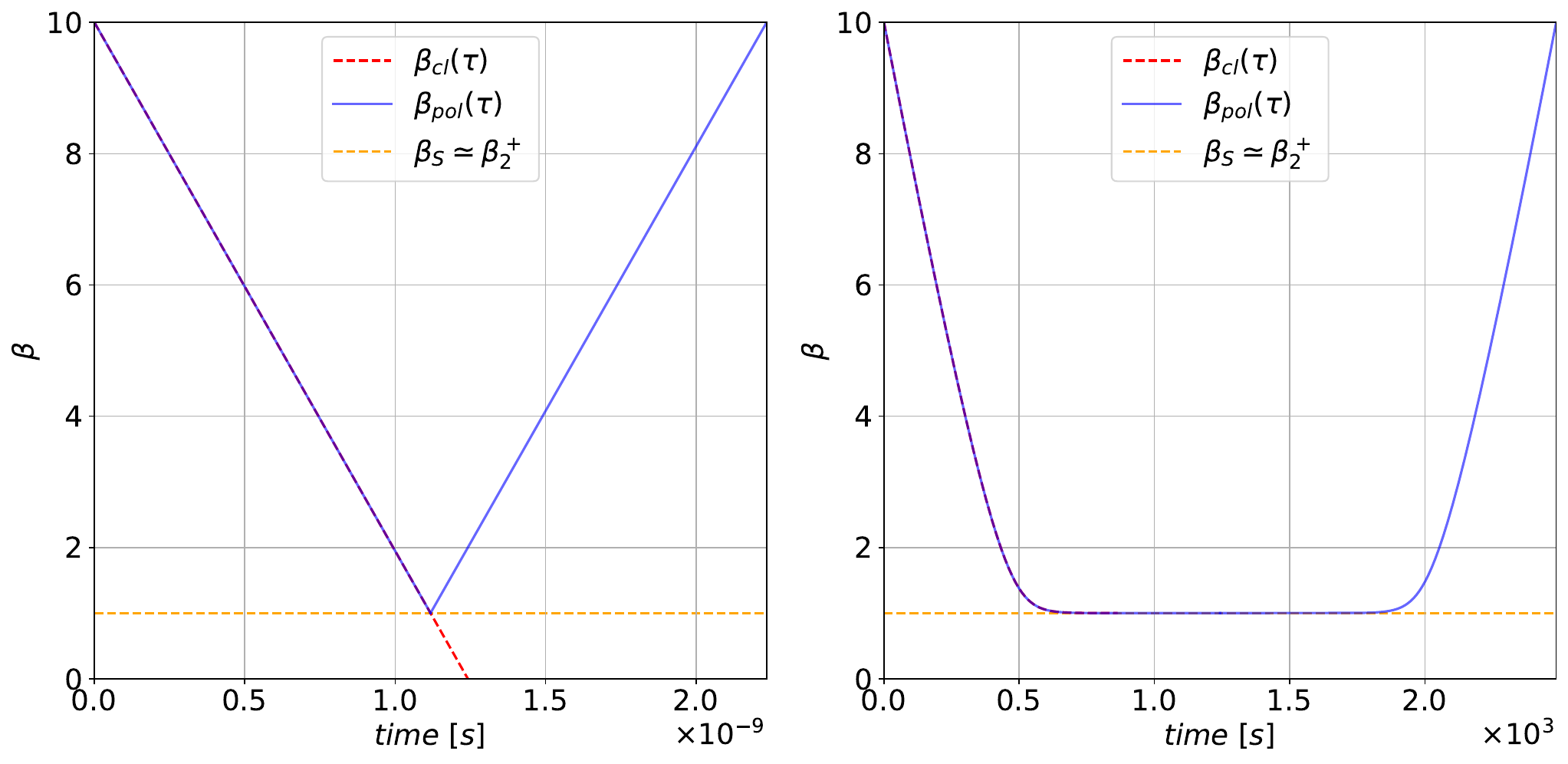}
    \caption{Polymer (blue) and classical (red) radial geodesics in $\tau$ (left) and $t$ (right) for a proton in-falling on $Sgr \ A^*$ with $E_{OMG} = (3.2 \pm 0.9) \cdot 10^{20} eV$. In orange, the constant $\beta = \beta^+_2$ line.}
    \label{fig:radial time geodesics OMG}
\end{figure}
\begin{figure}[H]
    \centering
    \includegraphics[width=1.0\textwidth]{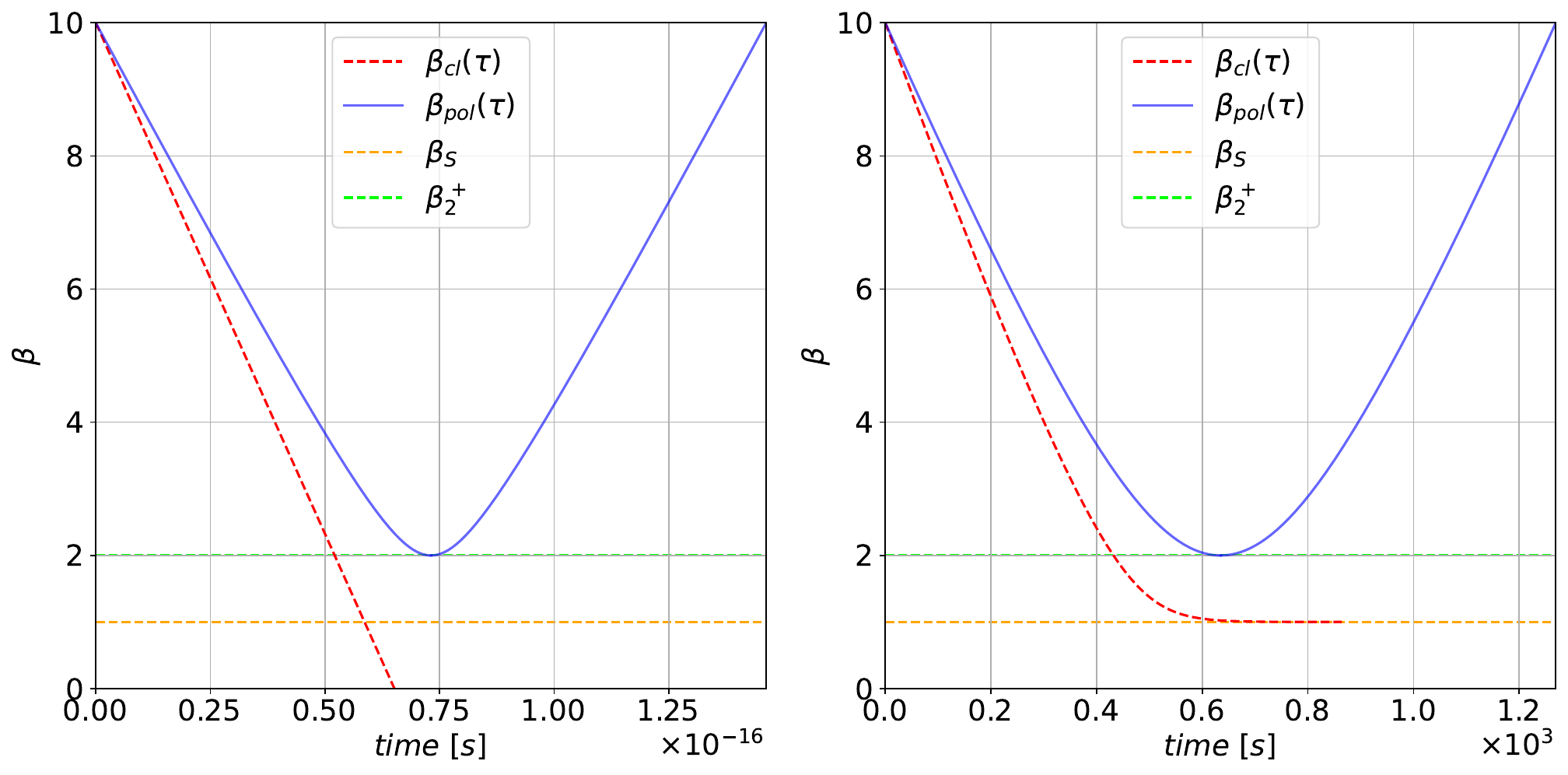}
    \caption{Polymer (blue) and classical (red) radial geodesics in $\tau$ (left) and $t$ (right) for a proton in-falling on $Sgr \ A^*$ with $E = 0.5 \hspace{1mm} E_P$. In orange, the constant $\beta = \beta_S$ line and in green the constant $\beta = \beta^+_2$ line.}
    \label{fig:radial time geodesics planck}
\end{figure}
\begin{figure}[H]
    \centering
    \includegraphics[width=1.0\textwidth]{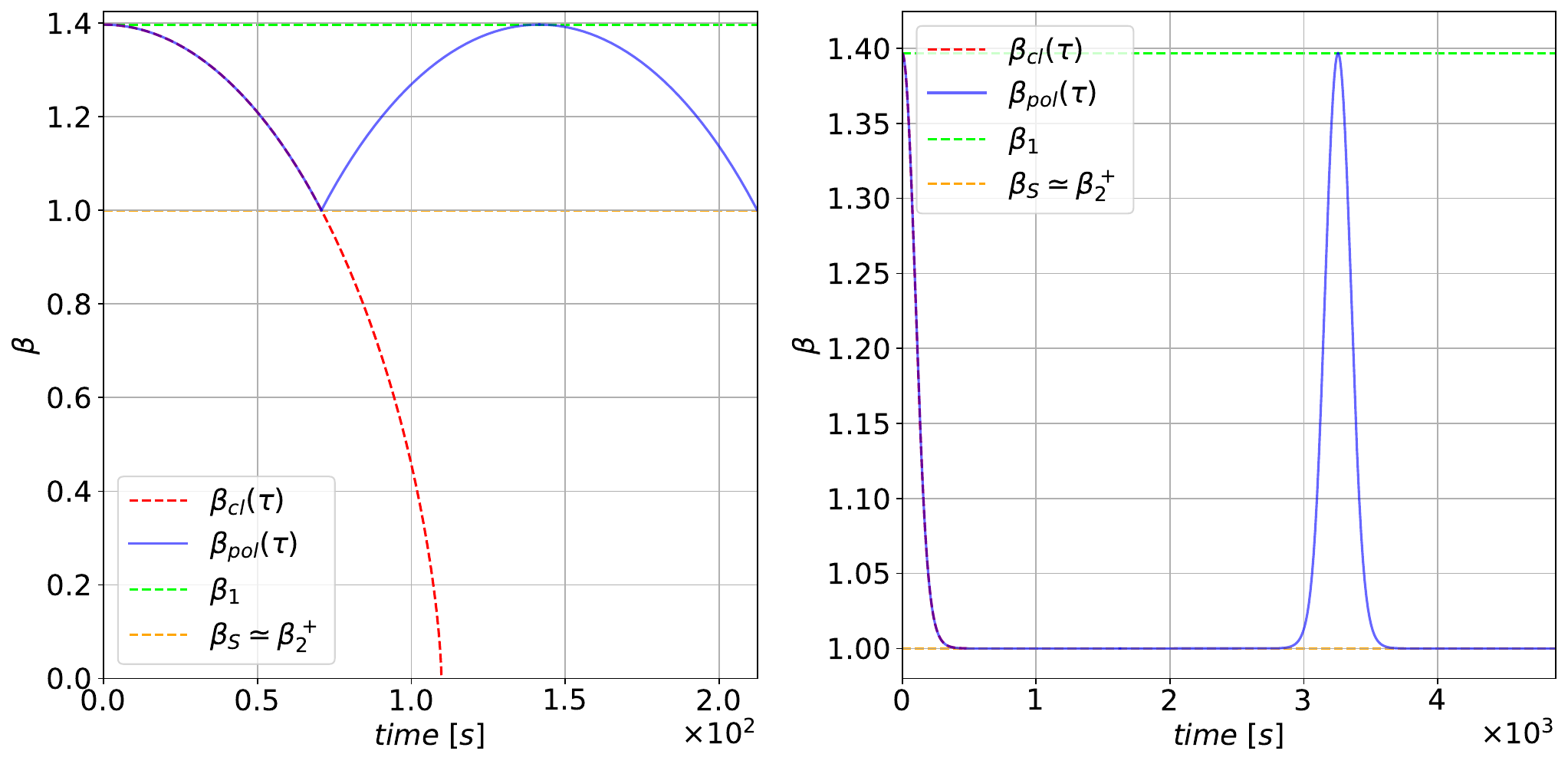}
    \caption{Polymer(blue) and classical (red) radial geodesics in $\tau$ (left) and $t$ (right) for a proton in-falling on $Sgr \ A^*$ with $E = 0.5 \hspace{1mm} GeV$. In orange, the constant $\beta = \beta^+_2$ line and in green the constant $\beta = \beta_1$ line.}
    \label{fig:radial time geodesics bound}
\end{figure}
It can be clearly seen that the two low energy cases, shown in figure \ref{fig:radial time geodesics OMG} and \ref{fig:radial time geodesics bound}, before the bounce are almost indistinguishable from the classical scenario, because it happens very close to the event horizon, namely $\beta^\pm_2 - \beta_S = 2 \cdot 10^{-16}$, so while it is immediately observable in proper time, in coordinate time the polymer geodesic flattens close to the horizon similarly to the classical one, but after a non-infinite amount of time the bounce can be observed.\\
In particular, the bound case has no classical counterpart and the particle remains confined in the region defined by the two solutions $\beta_2^+$ and $\beta_1$ following a periodic motion. It can be noted that we neglected any energy dissipation, which would bring the upper inversion point $\beta_1$ closer to the lower one $\beta_2^+$ with each bounce.\\
Instead, going to Planck scale energies, shown in figure \ref{fig:radial time geodesics planck}, the bounce happens much further away from the horizon, meaning that it is clearly and immediately observable in both reference frames, as the coordinate time geodesic does not flatten on the inversion point and the polymer scenario clearly diverges from the classical one.\\
Similar bouncing scenarios, in the context of radial dust collapse, were also found in the loop quantum gravity community\cite{Mnch2021, Boldorini_2024, Achour2020, Achour2020v1, BenAchour2020, Liu2014}, where the in-falling dust, that could be considered as a collection of single particles, bounces at different radii during gravitational collapse.\\
In particular, \cite{Boldorini_2024,Achour2020v1, BenAchour2020} found an impermeable event horizon, like we did here, under some specific conditions.

\section{Circular Orbits}
\label{sec:circ_pol}
\begin{figure}[H]
  \centering
  \begin{minipage}[b]{0.47\textwidth}
    \centering
    \includegraphics[width=\textwidth]{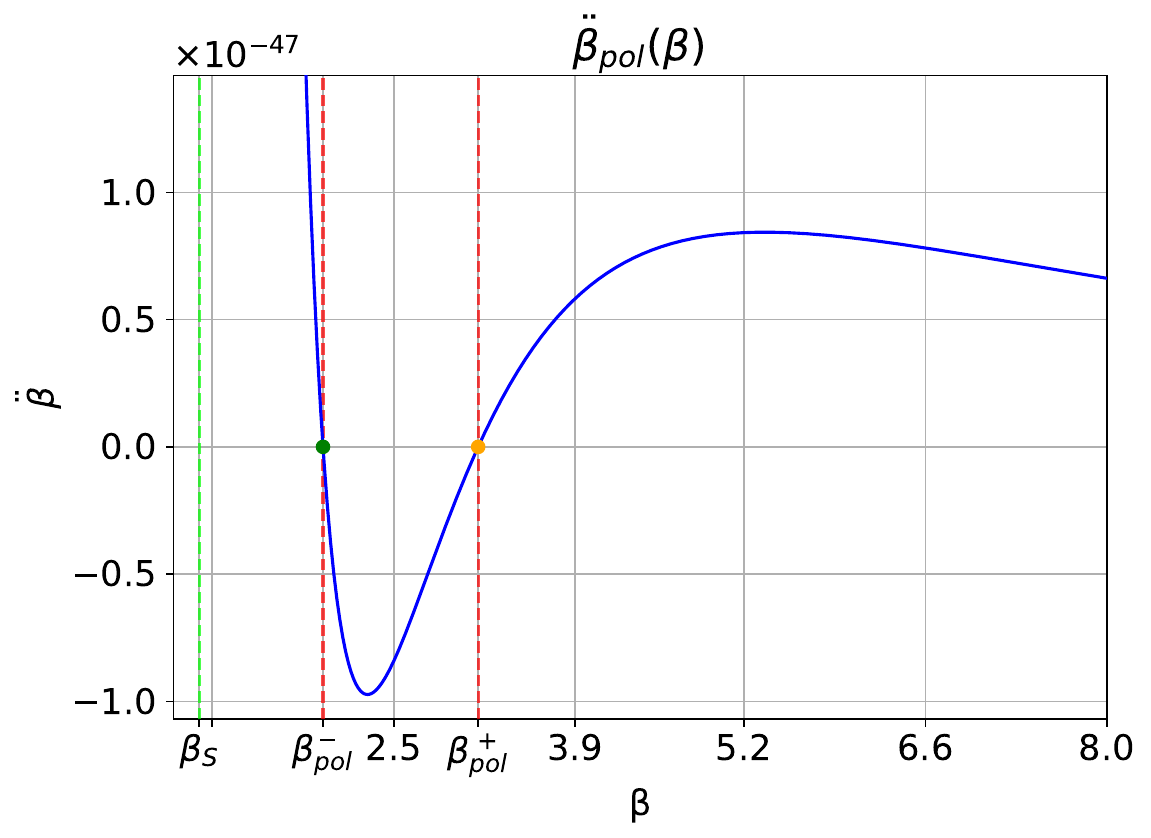}
  \end{minipage}
  \begin{minipage}[b]{0.47\textwidth}
    \centering
    \includegraphics[width=\textwidth]{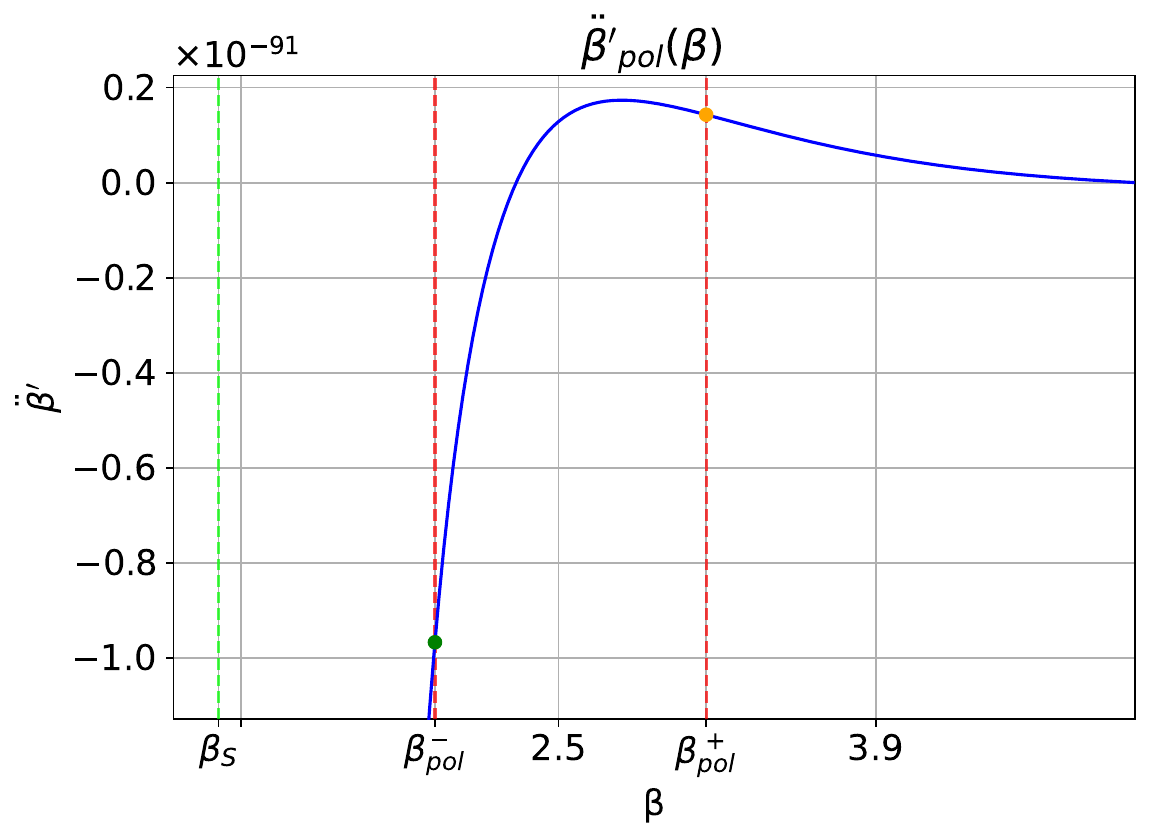}
  \end{minipage}
  \begin{minipage}[b]{0.47\textwidth}
    \centering
    \includegraphics[width=\textwidth]{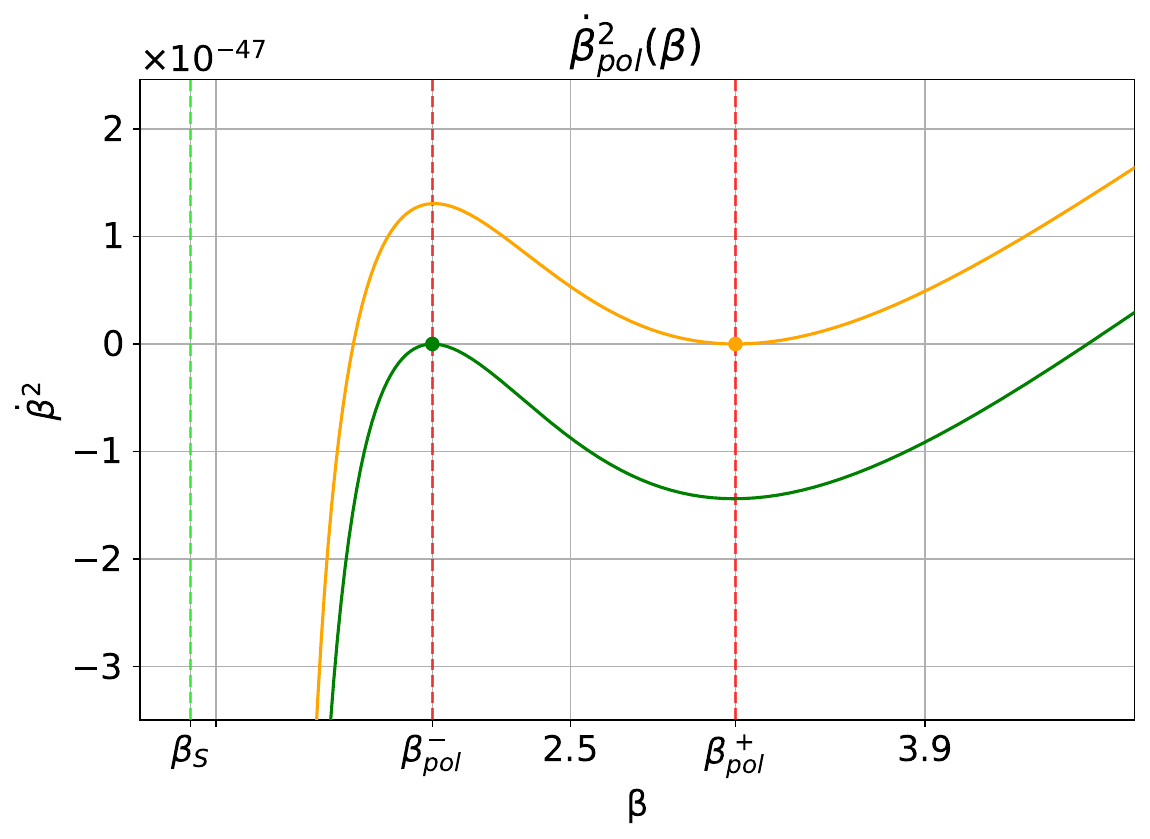}
\end{minipage}
    \begin{minipage}[b]{0.47\textwidth}
	\caption{Polymer circular orbits $\beta^\pm_{pol}$ outside the horizon for an electron orbiting $Sgr \ A^*$ with $\ell = 4.84 \cdot 10^{22}$, $\tilde E_2(\beta^+_{pol}) = 2.07 \cdot 10^{22}$ (orange plots) and $\tilde E_2(\beta^-_{pol}) = 2.09 \cdot 10^{22}$ (green plots). \vspace{30.7mm}}
    \label{fig:polymer out circular orbit}
    \end{minipage}
\end{figure}

\begin{figure}[H]
  \centering
  \begin{minipage}[b]{0.47\textwidth}
    \centering
    \includegraphics[width=\textwidth]{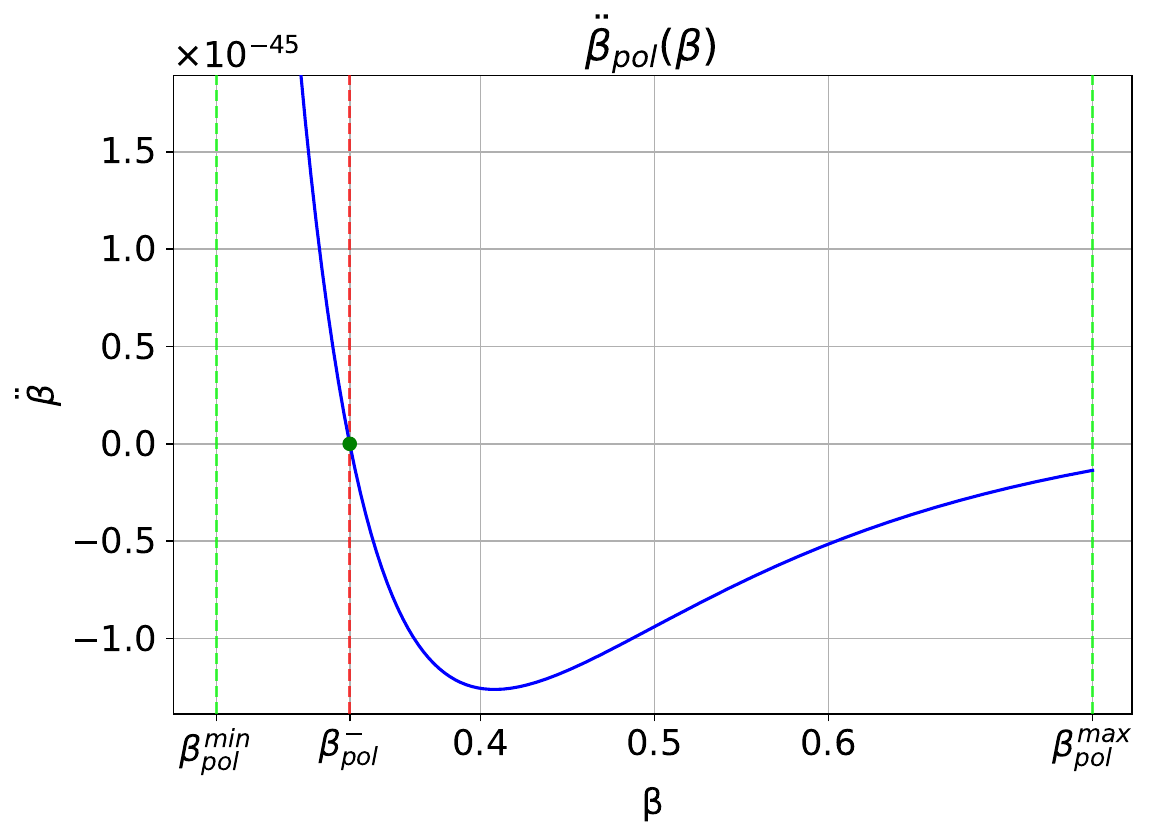}
  \end{minipage}
  \begin{minipage}[b]{0.47\textwidth}
    \centering
    \includegraphics[width=\textwidth]{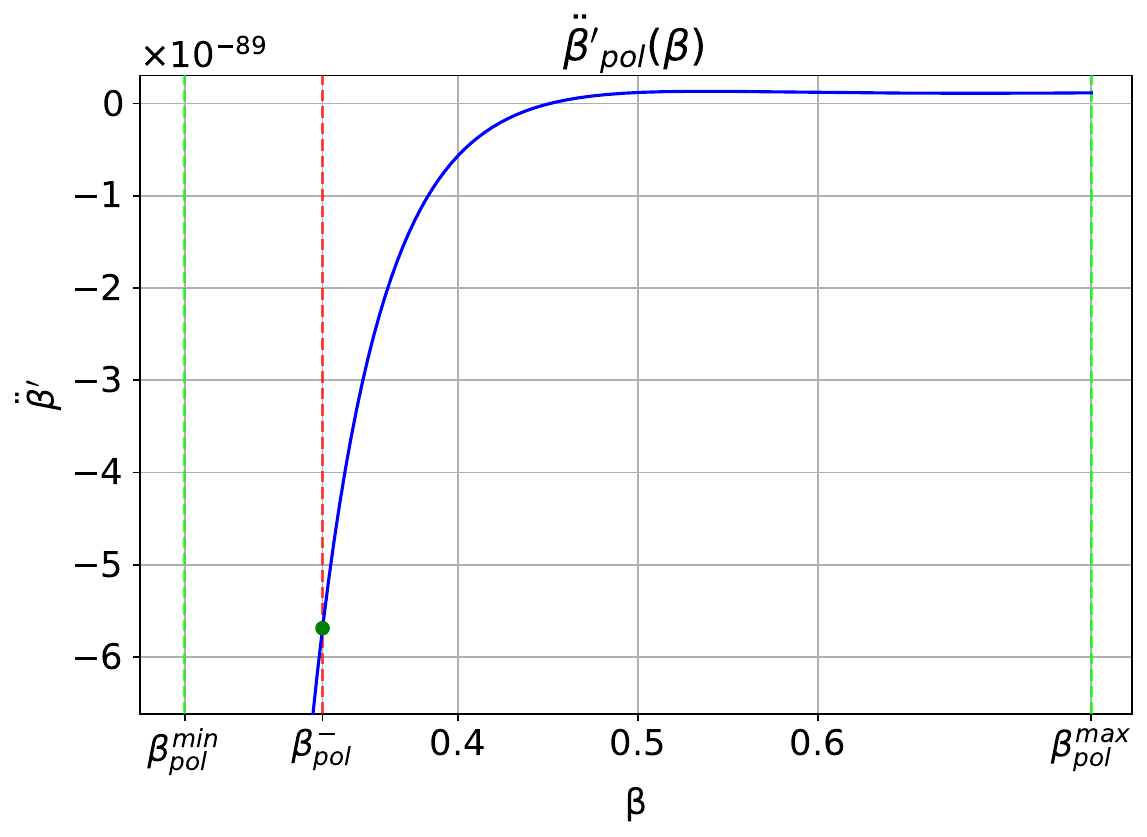}
  \end{minipage}
  \begin{minipage}[b]{0.47\textwidth}
    \centering
    \includegraphics[width=\textwidth]{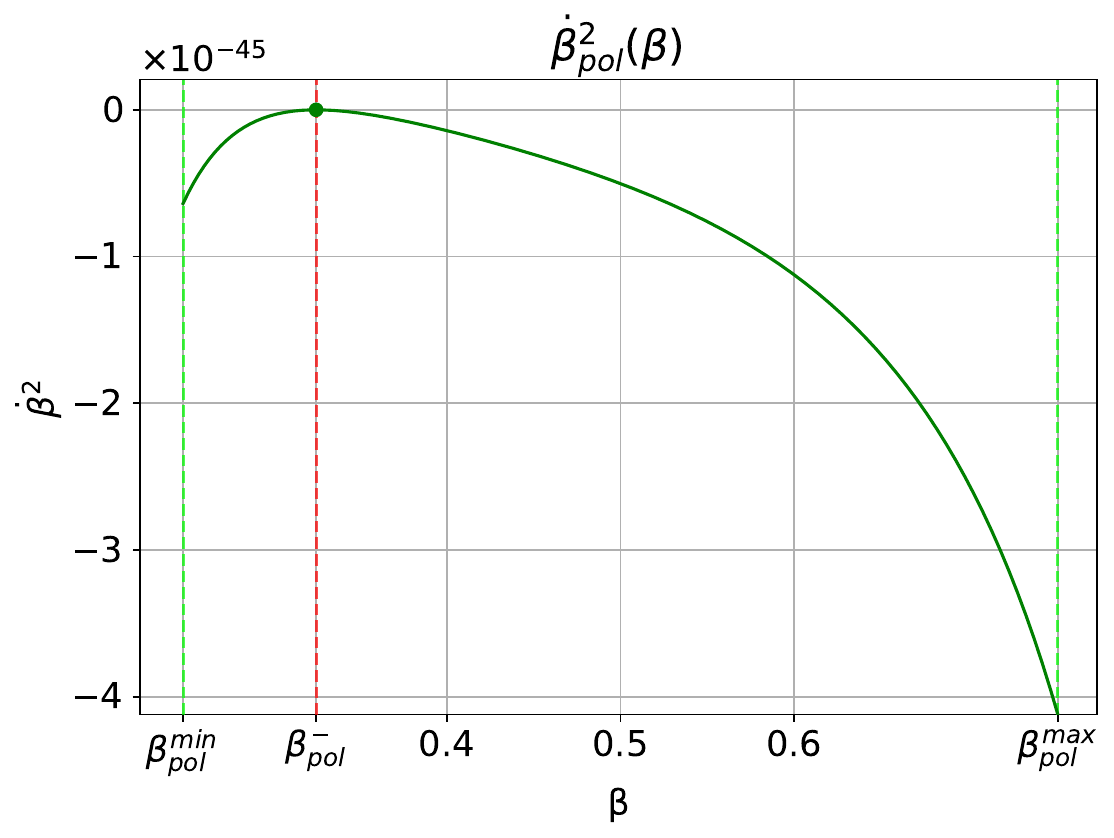}
  \end{minipage}
  \begin{minipage}[b]{0.47\textwidth}
    \caption{Polymer circular orbit $\beta^-_{pol}$ inside the horizon for an electron orbiting $Sgr \ A^*$ with $\ell = 1.03 \cdot 10^{22}$ and $\tilde E_2(\beta^-_{pol}) = 1.92 \cdot 10^{22}$. \vspace{37.2mm}}
    \label{fig:polymer in circular orbit}
  \end{minipage}
\end{figure}

To study circular orbits, we considered the full Hamiltonian given by equation (\ref{eqn:full_ham}), so that the Hamiltonian constraint now becomes:
\begin{equation}
	1 -\frac{\tilde E^2}{1-\frac{1}{\beta}} + \frac{\sin^2(\mu_0 p_r)}{\alpha^2}\left( 1-\frac{1}{\beta} \right) 
 + \frac{\ell^2}{\beta^2} = 0
	\label{eq:polymer general constraint}
\end{equation}
The polymer geodesics derived from the Hamilton equations are:
\begin{equation}
	\begin{cases}
		\dot{t} = \frac{\tilde E}{1-\frac{1}{\beta}} \\
		\dot{\beta} = \frac{\left(1-\frac{1}{\beta}\right)}{\alpha r_S} \sin (\mu_0 p_r) \cos (\mu_0 p_r) \\
		\dot{\varphi} = \frac{\ell}{r_S \beta^2}
	\end{cases}
	  ; \hspace{3mm}
	\begin{cases}
		\dot E = 0  \\
		\dot{p}_r = -\frac{m}{2 \ r_S} \left(\frac{\sin ^2(\mu_0 p_r)}{\alpha^2 \beta^2}+\frac{\tilde E^2 }{\beta^2 \left(1-\frac{1}{\beta}\right)^2} - \frac{2\ell^2}{\beta^3}\right) \\
		\dot \ell = 0 
	\end{cases}
	\label{eq:hamilton general equations var&mom}
\end{equation}
To study polymer circular orbits we solved the general system which defines circular orbits: 
\begin{equation}
\begin{cases}
\dot \beta = 0\\ 
\ddot \beta = 0
\end{cases}
\end{equation}
Where $\ddot \beta = \left[\dot \beta, H\right] = \dot \beta \frac{\partial \dot \beta}{\partial \beta} + \dot p_r \frac{\partial \dot \beta}{\partial p_r}$.\\
Solving $\dot \beta = 0$ is equivalent to solving $\dot \beta^2 = 0$, allowing us to remove the dependence on $p_r$ by substituting the sine squared from the Hamiltonian constraint in equation (\ref{eq:polymer general constraint}):
\begin{equation}
    \sin^2(\mu_0 p_r) = \frac{1}{1-\frac{1}{\beta}}\alpha^2 \left(\frac{\tilde E^2}{1-\frac{1}{\beta}}-1 - \frac{\ell^2}{\beta^2}\right)
\end{equation}
So the system becomes:
\begin{equation}
\begin{cases}
\dot \beta^2 = 0\\ 
\ddot \beta = 0
\end{cases} 
\qquad \Longrightarrow \qquad
    \begin{cases}
        T_1 \cdot T_2 = 0\\
         T_3 \cdot T_4 = 0
    \end{cases}
    \label{eq:circular system dimensionless}
\end{equation}
Where the four terms $T_i$ are given by:
\begin{align}
       T_1&= \left[ \ell^2 (-1 + \beta) + \beta^2 \left( -1 + \beta - \tilde{E}^2 \beta \right) \right]\\
        T_2 &=\left\{ \ell^2 \alpha^2 (-1 + \beta) + \beta + \beta^2 \left[ -2 + \beta + \alpha^2 \left( -1 + \beta - \tilde{E}^2 \beta \right) \right] \right\} \\
       T_3&=\left[ \ell^2 (-1 + \beta)(-1 + 2\beta) + \beta^2 (-1 + \beta - 2 \tilde{E}^2 \beta) \right]\\
       T_4 &= \left\{ 2 \ell^2 \alpha^2 (-1 + \beta) + \beta + \beta^2 \left[ -2 + \beta - 2 \alpha^2 \left( 1 + (-1 + \tilde{E}^2)\beta \right) \right] \right\}
    \label{eq:circular terms}
\end{align}
We solved the first equation with respect to the parameter $\tilde E^2$, which yields:
\begin{align}
       T_1 &= 0 \qquad\rightarrow\qquad \tilde E^2_1 = \frac{\ell^2 \beta + \beta^3 - \ell^2 - \beta^2}{\beta^3}\\
        T_2 &= 0 \qquad\rightarrow\qquad \tilde E^2_2 = \frac{
\beta^3 - 2 \beta^2 + \beta + \ell^2 \beta \alpha^2 + \beta^3 \alpha^2 - \ell^2 \alpha^2 - \beta^2 \alpha^2
}{
\beta^3 \alpha^2
}
    \label{eq:E squared constraints}
\end{align}
Where $\tilde E^2_1$ is associated to the classical Schwarzschild solutions shown in equation (\ref{eq:r pm classic}) and has the same existence conditions, while $\tilde E^2_2$ is associated with the purely polymer solutions.\\ 
The existence condition $\tilde E^2_2 \geq 0$ is verified in two intervals: if $\beta >1$ there are no existence conditions, while if $\beta <1$ the following conditions must hold:
\begin{equation}
\begin{cases}
        0< \ell < \frac{1}{2\alpha}\sqrt{\frac{1}{1+\alpha^2}}\\ \\
        \beta^{min}_{pol}  < \beta < \beta^{max}_{pol} < 1
\end{cases}
    \label{eq:E2 squared existence conditions}
\end{equation}
Where:
\begin{equation}
        \beta^{min}_{pol} = \frac{1 - \sqrt{1 - 4\alpha^2 \ell^2(1+\alpha^2)}}{2(1 + \alpha^2)}\qquad\qquad
        \beta^{max}_{pol} = \frac{1 + \sqrt{1 - 4\alpha^2 \ell^2 (1+\alpha^2)}}{2(1 + \alpha^2)}
    \label{eq:beta polymer min/max}
\end{equation}
This implies the astounding result that there could exist circular orbits inside the event horizon, thanks to the repulsive behavior of the polymer quantum representation which stabilizes them in a region characterized by space-like hypersurfaces.\\
\\
Substituting $E^2_2$ into $T_3\cdot T_4 = 0$ gives the polymer circular orbits:
\begin{equation}
    \beta^{\pm}_{pol} = \frac{1 + \ell^2 \alpha^2 \pm \sqrt{1 - 4\ell^2 \alpha^2 - 3\ell^2 \alpha^4 + \ell^4 \alpha^4}}{2 + \alpha^2}
\label{eq:beta pm polymer}
\end{equation}
These orbits exist in two regimes defined by their position with respect to the event horizon:\\
Outside the horizon ($\beta > 1$), we have  $\beta = \beta^\pm_{pol}$ and two cases: 
    \begin{align}
    \label{eq:out circular orbits 1}
       & \sqrt{\frac{4 + 3\alpha^2 + \sqrt{3(2 + \alpha^2)(2 + 3\alpha^2)}}{2\alpha^2}} \leq \ell \leq \frac{\sqrt{4 + 3\alpha^2}}{\alpha} \\  
       &\qquad \qquad \beta^-_{cl} < \beta^-_{pol} < \beta^+_{pol} \leq \beta_{cl}^{ISCO} \leq \beta^+_{cl}
  \end{align}
And:
    \begin{equation}
    \label{eq:out circular orbits 2}
        \ell > \frac{\sqrt{4 + 3\alpha^2}}{\alpha}   \qquad \beta^-_{cl} < \beta^-_{pol} < \beta_{cl}^{ISCO} < \beta^+_{pol} < \beta^+_{cl}
    \end{equation} 
Inside the horizon ($\beta < 1$) we have $\beta = \beta^-_{pol} $ and the following situation:
    \begin{equation}
        \ell < \frac{1}{2\alpha \sqrt{1+\alpha^2}} \qquad  \qquad \beta_{pol}^{min} < \beta^-_{pol} < \beta_{pol}^{max}
    \label{eq:in circular orbits}
    \end{equation}
To assess the stability of these orbits, we analyzed how the acceleration $\ddot \beta$ behaved around such extremal points by studying the sign of the function $\ddot \beta' = \frac{d\ddot \beta}{r_Sd\beta}$ to understand this behaviour.\\
Starting from the equilibrium point defined by (\ref{eq:circular system dimensionless}), if the acceleration becomes negative going to lower betas and positive going to bigger ones, the point is unstable because a perturbation $\pm|\delta\beta|$ pushes the particle away from the equilibrium point.
So $\ddot \beta$ is an increasing function with respect to $\beta$ and $\ddot \beta'(\beta_{eq})>0$.\\
Vice versa, if the acceleration becomes positive going to lower betas and negative going to bigger ones, the point is stable because a perturbation $\pm|\delta\beta|$ will make it return to the equilibrium point.
So $\ddot \beta$ is a decreasing function with respect to $\beta$ and $\ddot \beta'(\beta_{eq})<0$.\\
From this stability study we discovered that $\beta^+_{pol}$ is the unstable orbit, while $\beta^-_{pol}$ is the stable orbit.\\
We can note that with a Planckian $\mu_0$, the conditions shown in the equations (\ref{eq:out circular orbits 1}) and (\ref{eq:out circular orbits 2}) to have polymer circular orbits outside the event horizon is satisfied only in the Planck Energy regime since the leading order term is $\frac{1}{\alpha} = \frac{1}{\mu_0 m}$.\\
Figures \ref{fig:polymer out circular orbit} and \ref{fig:polymer in circular orbit} show the functions $\ddot \beta$, $\ddot \beta'$ and $\dot\beta^2$ in the case of an electron orbiting $Sgr \ A^*$ for the orbits both outside and inside the horizon respectively, with Planckian $\mu_0$.
First of all we note that the polymer effective model reproduces the classical set of circular orbits, alongside with a new distinct set of orbits induced by the polymer corrections. This new set of orbits can exist only in the Planckian energy regime, which imposes that the stable polymer orbit is always beneath the classical ISCO.\\
Furthermore, taking the $\ell \to +\infty$ limit, it can be seen that the polymer ISCO coincides with the classical IUCO $\beta = \frac{3}{2}$.\\
It can be noticed that the repulsive behavior found in the radial scenario is present also in the circular case, allowing stable circular orbits in the region bounded by the classical IUCO and ISCO, which would be a clear sign of the polymer effects, if observed.\\
In \cite{Saadati:2023jym,Ye:2023qks, Chen:2024sbc, Huang:2025vpi,Du:2024ujg}, where both massless and massive circular orbits around Loop Quantum Gravity- inspired polymerized black holes where analyzed, similar results where found. In particular, in \cite{Saadati:2023jym,Chen:2024sbc, Du:2024ujg} it was also found that the ISCO is reduced due to quantum gravity effects, while \cite{Ye:2023qks} proved that also the photon ring shrinks due to quantum effects, demonstrating a common trend in quantum gravity modified geodesics.\\
In \cite{Huang:2025vpi} was found instead that the ISCO for holonomy corrected black holes is the same as the classical one, just like we found in our first-branch solution. Particularly, they found a set of circular orbit undistinguishable from the classical scenario, exactly like we found for the first set of circular orbits.
\section{Conclusions}
\label{sec:concl}
In this paper we analyzed the effects of quantizing a particle in a diffeomorphism-invariant framework on a Schwarzschild geometry. Specifically, we studied the geodesic motion of said quantum particle and the way on which this motion differs from the classical case.\\
In section \ref{sec:hamil_pol} we introduced the polymer framework needed to build the Hamiltonian of the particle. With this Hamiltonian, in sections \ref{sec:rad_pol} and \ref{sec:circ_pol}, we derived the equations of motion that we solved to characterize radial trajectories and circular orbits respectively.\\
The main result of the radial study was the appearance of an impermeability of the event horizon through the introduction of a classically forbidden region, where $\dot{\beta}^2<0$, that always contains the horizon. On such surface the particles bounce off, when radially in-falling, in the same way as the elastic scattering off an infinite potential barrier. The forbidden region width shrinks to zero as the energy decreases, with a Planckian thickness for particles in the low energy regime.\\
This suggests that one should consider transmission and reflection through the barrier in order to take into account tunneling of low energy particles, making the event horizon not completely impermeable, allowing the particles to cross it. We leave this phenomenon for further studies.\\ 
Furthermore, from the study of circular orbits, we discovered that the polymer quantization introduces a new set of solutions alongside the classical ones, with some striking features.\\
We have a lower limit for $\ell$, given by equation (\ref{eq:out circular orbits 1}), for which circular orbits are situated above the event horizon. Particularly, the stable orbits are always located between the classical ISCO and IUCO, making those orbits a distinctive sign of the polymer modified theory, even if such effects are present only in the Planck regime.\\
For every other value of $\ell$ below such limit, the polymer orbits are inside the event horizon and stable. This poses a conceptual problem since circular orbits inside the event horizon correspond to closed time-like curves, so the meaning of such solutions is not physically clear.\\
Further investigations must be performed to better clarify the meaning of these results.

\bibliography{biblio.bib}

\end{document}